\documentclass[reprint,twocolumn,showpacs]{revtex4-1}

\usepackage{graphicx,color}
\usepackage{indentfirst}
\usepackage{amsfonts}
\usepackage{amsmath}
\usepackage{amssymb}
\usepackage[T1]{fontenc}
\newcommand{\bq}{\begin{eqnarray}}
\newcommand{\eq}{\end{eqnarray}}
\newcommand{\be}{\begin{equation}}
\newcommand{\ee}{\end{equation}}

\newcommand{\tr}{\operatorname{tr}}
\newcommand{\ket}[1]{\left| #1 \right\rangle}

\newcommand{\de}[1]{\left( #1 \right)}
\newcommand{\De}[1]{\left[ #1 \right]}

\newcommand{\abs}[1]{\left| #1 \right|}

\newcommand{\eg}{{\it{e.g.~}}}
\newcommand{\ie}{{\it{i.e.~}}}
\newcommand{\etal}{{\it{et al.}\ }}

\definecolor{gray}{rgb}{.4,.4,.4}
\definecolor{deepgreen}{rgb}{.1,.6,.3}

\begin{document}

\title{Asymptotic Entanglement Dynamics Phase Diagrams for Two Electromagnetic Field Modes in a Cavity}

\author{R.~C.~Drumond$^{1,2}$}
\author{L.~A.~M.~Souza$^{1,3}$}
\author{M.~Terra~Cunha$^{4}$}

\affiliation{$^{1}$Departamento de F\'{\i}sica, Instituto de
Ci\^{e}ncias Exatas, Universidade Federal de Minas Gerais, CP 702,
CEP 30123-970, Belo Horizonte, Minas Gerais, Brazil}
\affiliation{$^{2}$Institute for Quantum Optics and Quantum
Information, Austrian Academy of Sciences, Boltzmanngasse 3, Vienna,
Austria} \affiliation{$^{3}$Central de Ensino e Desenvolvimento
Agr\'ario de Florestal, Universidade Federal de Vi\c{c}osa --
\emph{Campus} de Florestal, LMG 318 -- Km 6, CEP 35690-000,
Florestal, Minas Gerais, Brazil} \affiliation{$^{4}$Departamento de
Matem\'{a}tica, Instituto de Ci\^{e}ncias Exatas, Universidade
Federal de Minas Gerais, CP 702, CEP 30123-970, Belo Horizonte,
Minas Gerais, Brazil}

\begin{abstract}
We investigate theoretically an open dynamics for two modes of
electromagnetic field inside a microwave cavity. The dynamics is
Markovian and determined by two types of reservoirs: the ``natural''
reservoirs due to dissipation and temperature of the cavity, and an
engineered one, provided by a stream of atoms passing trough the
cavity, as devised by Pielawa \emph{et al.} [\emph{Phys. Rev. Lett.}
\textbf{98}, 240401 (2007)]. We found that, depending on the
reservoir parameters, the system can have distinct ``phases'' for
the asymptotic entanglement dynamics: it can disentangle at finite
time or it can have persistent entanglement for large times, with
the transition between them characterized by the possibility of
asymptotical disentanglement. Incidentally, we also discuss the
effects of dissipation on the scheme proposed in the above reference
for generation of entangled states.
\end{abstract}


\maketitle

\section{Introduction}

In recent years, the problem of entanglement dynamics has gained the
attention of the quantum-information community~\cite{esd}. Despite
the fact that some recent results contradict the intuition that
``the more entangled the better''~\cite{Bacon-Phys}, to protect
entanglement is yet an important concern when one wants to make
quantum computers. The first works on the ``fate'' of entanglement
on an open system focused on two or more qubit systems. Curious
names such as \emph{entanglement sudden death} have appeared, first
by the recognition of the possibility of entanglement between two
qubits to vanish in finite time, although coherences only decay
exponentially. Similar phenomena also happen for continuous
variables (CV) systems~\cite{esdcv}. A general geometrical picture
has already been offered showing that the long term behavior of
entanglement depends essentially on the set of asymptotic
states~\cite{GeoSudDeath}.

In a series of papers, Paz and Roncaglia studied some phase diagrams
for the asymptotic behavior of entanglement on two-mode Gaussian
states (GS) exposed to a common thermal environment \cite{PR}. In
particular, they built them in terms of squeezing and temperature
for two harmonic oscillators. For the resonant case, they found
three very distinct possible fates: entanglement can suffer from
sudden death, can enter in a perpetual cycle of death and birth, or
can be persistent. From a different perspective, Pielawa \etal made
a proposal of how to use an engineered reservoir to create two-mode
Gaussian entangled states in cavity quantum electrodynamics (CQED)
\cite{davidov}.

In this paper we make a threefold study: we generalize the studies
from Paz and Roncaglia, by constructing phase diagrams in terms of
different variables, for systems subjected to not only natural
reservoirs, but also this engineered one; we show how those phase
diagrams can be experimentally obtained; and incidentally we study
the robustness, against thermal noise, of the engineered reservoir
strategy for generating two-mode entangled states.

The next section is devoted to reviewing some basics about two-mode
GS and the proposal for generating entangled ones in
CQED~\cite{davidov}. Section \ref{sec:EDER} is the central part of
our study, where the entanglement dynamics of two modes subjected to
thermal noise and the engineered reservoir is discussed. The
experimental proposal is focused in Sec. \ref{sec:ExpProp}, followed
by the study of the robustness of the method suggested in
Ref.~\cite{davidov}. Discussion and concluding remarks close the
paper.


\section{Preliminaries}\label{sec:davidov}

Here we will review some necessary definitions and tools in the
theory of entanglement of GS and also the scheme to implement the
engineered reservoir.

\subsection{Entanglement in Gaussian States}

Many of the protocols and techniques from the entanglement theory of
finite dimensional systems can be adapted to CV systems, usually
restricted to the set of GS. For instance, the protocols of
teleportation \cite{teleporte,teleportecv} and quantum key
distribution \cite{qkd,qkdcv} have their analogs for CV systems (and
may be more robust experimentally \cite{expqks}); the
Peres-Horodecki \cite{peres,pereshorodecki} separability criteria
can be applied, and is also necessary and sufficient for two-mode GS
\cite{simon}; and entanglement quantifiers such as negativity
\cite{neg}, logarithmic negativity \cite{logneg}, and entanglement
of formation \cite{enform} can be computed within the formalism of
symplectic geometry \cite{negcv,enformcv,adesso}.

GS are defined as those states whose Wigner characteristic function
is Gaussian, so they are completely described by its first and
second statistical momenta. First momenta, mean values, can be
locally changed by redefining the mode operators, so all
entanglement information is given by second momenta. Choosing only
two modes, with destruction operators
$\hat{a}_{j}=(\hat{x}_{i}+i\hat{p}_{i})/\sqrt{2}$, $i=1,2$, and
corresponding quadrature amplitudes $\hat{x}_{i},\hat{p}_{i}$, the
Wigner characteristic function of a state $\hat{\rho}$ is given by
$\chi(z_{1},z_{2})=\text{Tr}[\hat{\rho}\exp{(z_{1}\hat{a}_{1}-z_{1}^{*}\hat{a}_{1}^{\dagger}+z_{2}\hat{a}_{2}-z_{2}^{*}\hat{a}_{2}^{\dagger})}]$,
where $z_{i}$ are complex numbers. The state second momenta, on the
other hand, are well grouped under its covariance matrix (CM):
\begin{eqnarray}
V_{\hat{\rho}}=\left( \begin{array}{cccc}
n_{1}+\frac{1}{2} & m_{1} & m_{s} & m_{c}\\
m_{1}^{*} & n_{1}+\frac{1}{2} & m_{c}^{*} & m_{s}^{*}\\
m_{s}^{*} & m_{c} & n_{2}+\frac{1}{2} & m_{2}\\
m_{c}^{*} & m_{s} & m_{2}^{*} & n_{2}+\frac{1}{2}
\end{array} \right),
\end{eqnarray}
where $n_{j}=\langle \hat{a}^{\dagger}_{j}\hat{a}_{j}\rangle$,
$m_{j}=-\langle \hat{a}_{j}^{2}\rangle$, $m_{s}=-\langle
\hat{a}_{1}\hat{a}_{2}^{\dagger}\rangle$, $m_{c}=\langle
\hat{a}_{1}\hat{a}_{2}\rangle$, and $\langle \hat{\xi} \rangle$
denotes the quantum expectation value $\text{Tr}\hat{\xi}\hat{\rho}$
of an observable $\hat{\xi}$. The state is Gaussian (with null first
momenta) if and only if
$\chi(z_{1},z_{2})=\exp{(-\frac{1}{2}\textbf{z}^\dagger
V_{\hat{\rho}}\textbf{z})}$, where
$\textbf{z}^t=(z_{1},z_{1}^{*},z_{2},z_{2}^{*})$.

We can write the CV as a block matrix, as follows:
\begin{eqnarray} V_{\hat{\rho}}=\left( \begin{array}{cc}
V_1 & C\\
C^\dagger & V_2
\end{array} \right),
\end{eqnarray}
where $V_i$ is a $2 \times 2$ matrix related to the mode $i$, and
$C$ is a $2 \times 2$ matrix that gives the correlations (both
quantum and classical) between the modes. One should note that to
represent a quantum state, a CM should also obey the generalized
{Robertson-Schr\"odinger} uncertainty relations~\cite{adesso}. A CM
not obeying such relations is called \emph{nonphysical}.

\begin{subequations}\label{Simon}
Simon~\cite{simon} has shown that, as for two qubits, the
Peres-Horodecki criterion \cite{pereshorodecki} is decisive for
entanglement of two-mode GS, and it can be given in terms of the
quantity:
\begin{equation} \label{simonquantity}
S\de{V_{\hat{\rho}}} = I_1 I_2 + (1/4-|I_3|)^2 - I_4 - 1/4
(I_1+I_2),
\end{equation}
where $I_{1,2} = \det V_{1,2}$, $I_3 = \det C$, and $I_4 =
\tr\De{V_1 Z C Z V_2 Z C^\dagger Z}$ are invariants under local
unitary operations, with $Z=\text{diag}\{1,-1\}$. A GS is separable
if and only if
\begin{equation}
S\de{V_{\hat{\rho}}} \geq 0,
\end{equation}
\end{subequations}
the so-called Simon criterion.

Entanglement in two-mode systems reminds Einstein-Podolsky-Rosen
(EPR) discussion on completeness of quantum mechanics~\cite{EPR}.
Indeed, entanglement in this system is closely related to squeezing
in ``EPR-like'' quadratures. Given any local operators
$\hat{Q}_{i},\hat{P}_{i}$ satisfying the commutation relations
$[\hat{Q}_{i},\hat{P}_{i}]=i\hbar$, for any arbitrary real number
$a\neq0$ one can define the pair of EPR-like operators
$\hat{u}=|a|\hat{Q}_{1}+\frac{1}{a}\hat{Q}_{2}$,
$\hat{v}=|a|\hat{P}_{1}-\frac{1}{a}\hat{P}_{2}$. Duan \emph{et al.}
\cite{DGCZ} have shown that if a state is separable, then $\langle
\Delta\hat{u}^{2}\rangle+\langle \Delta\hat{v}^{2}\rangle \geq
a^2+1/a^2$. Hence, if this pair of EPR operators is squeezed enough,
\emph{i.e.}, if the sum of their variances violates the inequality,
the state is entangled for sure.

For GS, this criterion is necessary and sufficient to decide
separability, which we shall call the Duan-Giedke-Cirac-Zoller
(DGCZ) criterion. This is done representing the CM in a standard
form, through local Gaussian operations, so that the validity of the
above inequality applied to the matrix in this form, with $a$
determined by its coefficients, implies that the state is separable.
Restricting further to the set of symmetric states, it is sufficient
to consider $|a|=1$, and the procedure amounts to finding, through
local rotations and squeezing of the quadratures
$\hat{x}_{i},\hat{p}_{i}$, which pair of EPR-like operators have the
least value for the sum of their variances.

We shall deal mostly with symmetric GS (\ie $I_1 = I_2$), and for these we shall use
the Entanglement of Formation as an entanglement quantifier, which
has an explicit formula~\cite{rigol}
\begin{equation}\label{E_fRigol}
E_{F}(\rho)=f(2\sqrt{I_{1}+|I_{3}|-\sqrt{I_{4}+2I_{1}|I_{3}|}})
\end{equation}
where $f(x)=c_{+}(x)\log_{2}c_{+}(x)-c_{-}(x)\log_{2}c_{-}(x)$, with
$c_{\pm}(x)=\frac{1}{4}(x^{-1/2}\pm x^{1/2})^{2}$.

\subsection{Two-mode entanglement from engineered reservoir}

To make the context clear, we now review the scheme for
constructing the common squeezing reservoir between the modes \cite{davidov}.
We also take the opportunity to introduce notation and to explicit
 the dependence of the
final master equation on the several parameters of the setting.

Consider two modes of electromagnetic field of a high quality
microwave cavity, with frequencies $\omega_{1}$ and $\omega_{2}$. As
before, $\hat{a}_i$ denotes the annihilation operator for mode $i$. The
engineered reservoir is provided by a stream of atoms passing
through the cavity. The atoms are first prepared in a specific
superposition of two Rydberg states denoted $\ket{g}$ and $\ket{e}$,
then they pass through the cavity, where they can interact with the
two nondegenerate modes, while a classical field (injected
externally in the setup of open cavities) saturates the dipole
transition, pumping the two modes. The Hamiltonian that describes
such setup is:
\begin{eqnarray}
\hat{H} &=& \hbar \omega_0 \hat{\sigma}^+ \hat{\sigma}^- + \hbar
\Omega (e^{- i \omega_L t} \hat{\sigma}^+ + e^{i \omega_L t}
\hat{\sigma}^-) \nonumber
\\ \label{hamilton}&& + \sum_i [\hbar \omega_i
\hat{a}_i^\dagger \hat{a}_i + \hbar g_i (\hat{a}_i
\hat{\sigma}^+ + \hat{a}_i^\dagger \hat{\sigma}^-)], \label{Hamilton}
\end{eqnarray}
where $\omega_0$ is the transition frequency between the atomic
levels, $g_i$ are the coupling constants between the atom and the
modes, $\hat{\sigma}^+$ and $\hat{\sigma}^-$ are the atomic
ladder operators. The coupling with the external classical field,
with strength $\Omega$, is described by the time dependent part. For
future reference, we set $\Delta = \omega_L - \omega_0$, the
detuning between the classical field and the atomic level.

The authors explore different approximations and mode redefinitions.
Here we will be concerned with the regime when (i) the atomic
coupling with the classical field is  much stronger than with
the cavity modes: $\abs{\Omega} \gg \abs{g_i}$; (ii) defining $d =
\sqrt{\Delta^2 + 4 \Omega^2}$, and choosing $\omega_L$ obeying
$\omega_L - \omega_1 = \omega_2 - \omega_L = d$, with the condition
$g = g_1 = g_2$. Under this regime the interaction Hamiltonian can
be approximated by:
\begin{subequations}\label{Hint}
\begin{eqnarray}
\hat{H}_{int}&\simeq& - \hbar \Omega_b (\hat{b}_1 \hat{\pi}^- +
\hat{b}_1^\dagger \hat{\pi}^+), \text{~~~~~ if $\Delta > 0$},  \\
\hat{H}_{int}&\simeq& \hbar \Omega_b (\hat{b}_2^\dagger \hat{\pi}^-
+ \hat{b}_2 \hat{\pi}^+), \text{~~~~~~~ if $\Delta < 0$},
\end{eqnarray}
\end{subequations}
where $\hat{\pi}^+$ and $\hat{\pi}^-$ are ladder operators for
semiclassical dressed states $\ket{+}=\sin \theta \ket{g} + \cos
\theta \ket{e}$ and $\ket{-}=\cos \theta \ket{g} - \sin \theta
\ket{e}$, with $\tan \theta = 2 \Omega / (d - \Delta)$, $\Omega_b$
is related to the coupling constant between the atoms and the cavity
modes, $\Omega_b = g \sqrt{(1-\mu)/(1+\mu)}$, where $\mu = \tan^2
\theta$ [$\mu = (\tan \theta)^{-2}$] if $|\tan \theta| < 1$ [$|\tan
\theta|>1$], \emph{i.e.}, $\mu$ is determined by the classical field
parameters. The new modes are $\hat{b}_{1(2)} =
\hat{S}^\dagger(r_\mu) \hat{a}_{1(2)}
\hat{S}(r_\mu)=\cosh{|r_{\mu}|}a_{1(2)}-\frac{r_{\mu}}{|r_{\mu}|}\sinh{|r_{\mu}|}a^{\dagger}_{2(1)}$
defined by the well-known two-mode squeezing operator: $
\hat{S}(r_\mu) = \exp (r_\mu^* \hat{a}_1 \hat{a}_2 - r_\mu
\hat{a}_1^\dagger \hat{a}_2^\dagger)$, and $r_\mu$ is the squeezing
parameter $r_\mu = \text{arctanh} \mu$.

If after the interaction time the atoms are ignored, the Hamiltonian
of Eq.~\eqref{Hamilton} implies an open system effective dynamics
for the field modes. Eqs.~\eqref{Hint} show that for $\Delta > 0$
($\Delta < 0$), the interaction reduces to a (anti-) Jaynes-Cummings
between the classically dressed atom and mode $\hat{b}_{1(2)}$. For
$\Delta > 0$ ($\Delta < 0$), one simulates null temperature
dissipation in mode $\hat{b}_{1(2)}$ if one prepares atoms in the
state $\ket{+}$ ($\ket{-}$) and allow interaction for times, $\tau$,
obeying $\Omega_b \tau\ll 1$. We shall call these type 1 (2) atoms.
If a stream of type $j$ atoms passes trough the cavity, one at a
time, the field dynamics will be Markovian, given by a differential
equation in Lindblad form~\cite{lindblad}:
\begin{subequations}
\begin{equation}\label{master1}
\frac{d\hat{\rho}}{dt}=\mathcal{D}_{j,Eng}(\hat{\rho}) ,
\end{equation}
where the effective engineered dissipator is given by~\cite{morigi}
\begin{equation}\label{Dengj}
\mathcal{D}_{j,Eng}(\hat{\rho})= 2\kappa_{j} (2 \hat{b}_j \hat{\rho}
\hat{b}_j^\dagger - \hat{b}_j^\dagger \hat{b}_j \hat{\rho} -
\hat{\rho} \hat{b}_j^\dagger \hat{b}_j),
\end{equation}
while $\kappa_{j} = (r_{at,j} \Omega_b^2 \tau^2)/4$ with $r_{at,j}$
the atomic preparation rate.
\end{subequations}

We note that, at first sight, the large frequency separation between
the modes would forbid, in the regime considered here, the presence
of combined terms between the modes, such as $a_{1}a_{2}^{\dagger}$
in the master equation, since they are fast oscillating compared to
the total time scale of the experiment and even the interaction time
of each atom [which is essential, in order to use the approximated
Hamiltonian \eqref{Hint}]. Though, the crucial point is that the
terms $a_{1}^{(\dagger)}a_{2}^{(\dagger)}$, the only combined ones
present in the master equation, do not oscillate in the laser
reference frame. We will come back to this point later when
discussing the state evolution.

\section{Entanglement Dynamics under Engineered and Thermal Reservoirs}\label{sec:EDER}

If a random source defines together the type of the atom and the
suitable DC electrical field in the cavity,  the dissipator for the
engineered reservoir will acquire the form:
\begin{equation}\label{DengT}
\mathcal{D}_{Eng}(\hat{\rho})=\mathcal{D}_{1,Eng}(\hat{\rho})+\mathcal{D}_{2,Eng}(\hat{\rho}),
\end{equation}
with $\kappa_j=(r_{at,j} \Omega_b^2 \tau^2)/4$, $r_{at,j}$ being the
type $j$ atoms flux.

The entanglement behavior will get richer when considered together with the
natural dissipation and thermal noise on the modes, effects that, in
usual experiments with microwave cavities, can be well described by
a Lindblad equation with dissipator $\mathcal{D}_{Nat}$ given by:
\begin{align}\label{Dnat}
\mathcal{D}_{Nat}(\hat{\rho})&=& \sum_i \lambda_i (n_{T_i} + 1) (2
\hat{a}_i \hat{\rho} \hat{a}_i^\dagger -
\hat{a}_i^\dagger \hat{a}_i \hat{\rho} - \hat{\rho} \hat{a}_i^\dagger \hat{a}_i)\nonumber \\
&& + \lambda_i n_{T_i} (2 \hat{a}_i^\dagger \hat{\rho} \hat{a}_i -
\hat{a}_i \hat{a}_i^\dagger \hat{\rho} - \hat{\rho} \hat{a}_i
\hat{a}_i^\dagger),
\end{align}
where $n_{T_i}$ denotes the number of thermal photons and
$\lambda_{i}$ the decay rate for mode $i$. Since the number of
thermal photons is approximately the same for both modes if
$\abs{\omega_1 - \omega_2} \ll \omega_i$, we shall assume from now
on $n_{T_i} = n_T$.

At last, the full dynamics of the system will be described by the
Lindblad equation:
\begin{equation}\label{master}
\frac{d\hat{\rho}}{dt}=\mathcal{D}_{Nat}(\hat{\rho})+\mathcal{D}_{Eng}(\hat{\rho})
,
\end{equation}
with the dissipator given by Eqs.~\eqref{DengT} and \eqref{Dnat}.

In an experimental setting one should actually consider the
evolution in the laser reference frame. In this case, fast
oscillations (with frequencies of the order $d$) between the modes
will take place. As a consequence, in the course-grained scale
necessary for the approximations to be valid, some coherences of the
density matrix will vanish, making the theoretical analysis more
complicated. Nevertheless, the above equation will still be valid
for some initial states, i.e., they are not affected by these
oscillations. Hence, in the rest of this section we shall explore
the theoretical properties of Eq.~\eqref{master}, considering
several initial states, but later, in the experimental proposal, we
must guarantee the use of such robust states against this
coarse-graining effect.

\subsection{Symmetric engineered reservoir}

Now we pass to the study of entanglement dynamics under
Eq.~\eqref{master}. Given the specific form of the dissipators,
Gaussianity is preserved, so we are safe to restrict ourselves to
GS. Let us suppose first that both types of atoms enter in the
cavity with equal probability, so that
$\kappa_{1}=\kappa_{2}=\kappa$, and also, for the sake of
simplicity, that $\lambda_{1}=\lambda_{2}=\lambda$. The equations of
motion for the second momenta in this case become:
\begin{subequations} \label{CMDyn}
\begin{eqnarray}
 \dot{n_{j}}&=&-2(\kappa+\lambda)n_{j}+2\kappa|B|^{2}+2\lambda n_{T},\\
 \dot{m_{j}}&=&-2(\kappa+\lambda)m_{j},\\
 \dot{m_c}&=&-2(\kappa+\lambda)m_c+2\kappa AB^{*}, \\
 \dot{m_s}&=&-2(\kappa+\lambda)m_s,
\end{eqnarray}
\end{subequations}
with  $A = \cosh (r)$,  $B = e^{i \phi} \sinh (r)$, where $r$ is the
squeezing parameter and $\phi$ the squeezing angle in the definition
of modes $\hat{b}_j$ ($r_{\mu}=re^{i\phi}$). This gives a relaxing
dynamics with the asymptotic CM:
\begin{eqnarray}
V_{\hat{\rho}_{f}}=\left( \begin{array}{cccc}
n_{1,f}+\frac{1}{2} & 0 & 0 & m_{c,f}\\
0 & n_{1,f}+\frac{1}{2} & m_{c,f}^{*} & 0\\
0 & m_{c,f} & n_{2,f}+\frac{1}{2} & 0\\
m_{c,f}^{*} & 0 & 0 & n_{2,f}+\frac{1}{2}
\end{array} \right),
\end{eqnarray}
where $n_{1,f}=n_{2,f}=\frac{|B|^{2}+n_{T}R}{1+R}$,
$m_{c,f}=\frac{AB^*}{1+R}$, and the ratio $R=\lambda/\kappa$ was
introduced.

Simon criterion can now be applied to determine whether such CM
represents entangled or separable states. In the same reasoning as
in Ref.~\cite{GeoSudDeath}, $S\de{V_{\rho_f}}>0$ implies a deep
separable asymptotic state, \emph{i.e.,} a state belonging to the
interior of the separable states set. This can be seen noting that
$S$ is a continuous function, so there must exist a ``ball'' of GS
around the asymptotic one such that $S$ is strictly positive, that
is, a ball of separable GS. Hence, this translates dynamically as
sudden death of entanglement (SDE), because we can say for sure
that, for every initial GS, there will be a time $T$ such that the
entanglement will be zero for $t > T$. In principle, though, it
could undergo cycles before it vanishes for good, or an initially
separable state can acquire some entanglement and (necessarily) lose
it after some time. These ``non-asymptotic'' behaviors will be
discussed in more detail on Sec. \ref{Sub:Nonasymp}.

On the other hand, states satisfying $S\de{V_{\rho_f}}<0$ represent
a situation of (asymptotic) persistent entanglement (PE). Note that
entanglement can be created by the common reservoir, since this is
exactly the idea of the Ref.~\cite{davidov} proposal. But again, the
intermediate dynamics can exhibit richer features; for instance, an
initially entangled state can lose all entanglement at finite time
but (necessarily) recover it at later time, as will be exemplified
at Sec. \ref{Sub:Nonasymp}.

The exceptional situation is given by $S\de{V_{\rho_f}}=0$, when
each initially entangled state can show one of two fates: sudden
death of entanglement or asymptotic death of entanglement, depending
on the initial state. Contrary to the former two situations, this
one requires the knowledge of the whole dynamics in order to
determine which fate will occur.

The situation here studied allows only one asymptotic state for each
set of fixed parameters, that is why one can not see infinite cycles
of birth and death as in Ref.~\cite{PR}, which would appear for
dynamics with asymptotic periodic orbits, instead of asymptotic
states.

\begin{figure}[htp]
  \centering
  \includegraphics[width=8 cm]{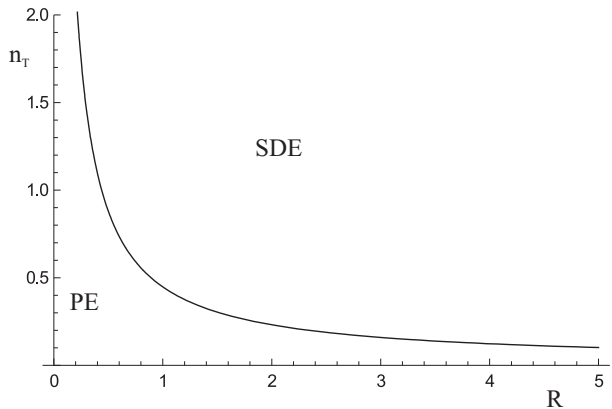}\\
  \caption{Representation of the asymptotic states determined by reservoir parameters $n_{T}$ and $R$ with respect to their entanglement, for $r=1$ and arbitrary $\phi$. Below the curve, they are entangled, and above, separable. This also determines the asymptotic properties for the entanglement dynamics of any initial state: for points above the curve, every initially entangled state exhibits SDE; for points below the curve, every initial state has some entanglement for large times; and for points exactly in the curve, entanglement can die at finite time or asymptotically.}\label{curva}
\end{figure}

In Fig.~\ref{curva}, we plot the regions in the parameter space $R
\times n_T $ where the asymptotic state is separable or entangled,
defining the fate of entanglement.  The boundary curve separating
the two regions in the diagram has the simple form:
\begin{equation}\label{eqcurva}
n_{T}=\frac{e^{2r}-1}{2R}.
\end{equation}
The physical interpretation is simple and meaningful. Given the
engineered reservoir, for any given positive coupling ratio, $R$,
there is a positive temperature, $n_T$, obeying \eqref{eqcurva} such
that below this temperature, the asymptotic state is entangled, due
to the common reservoir, while for temperatures above that critical
value, the asymptotic state is separable, when (local) thermal noise
prevails.

For any set of reservoir parameters, we applied the method of Duan,
Giedke, Cirac, and Zoller~\cite{DGCZ} and found the EPR-like
operators with least sum of variances, which gives
$\hat{X}_{1,\phi}-\hat{X}_{2,\phi}$ and
$\hat{P}_{1,\phi}+\hat{P}_{2,\phi}$, where
$(\hat{X}_{i,\phi},\hat{P}_{i,\phi})^T=\mathcal{R}_{2\phi}(\hat{x}_{i},\hat{p}_{i})^T$,
$\mathcal{R}_{2\phi}$ is the matrix representing a rotation in the
plane trough an angle $2\phi$, $\phi$ being the squeezing angle
defined by the reservoir. This is expected, since the engineered
reservoir tries to lead the initial state to a usual two-mode
squeezed state, known to be squeezed in these quadratures. The
natural reservoir, on the other hand, enlarges their spreading. So
the final decision of whether the asymptotic state will be separable
or entangled will depend on this competition between the reservoirs:
one trying to squeeze the collective quadratures, the other trying
to spread them.

\begin{figure}[htp]
  \centering
  \includegraphics[width=8 cm]{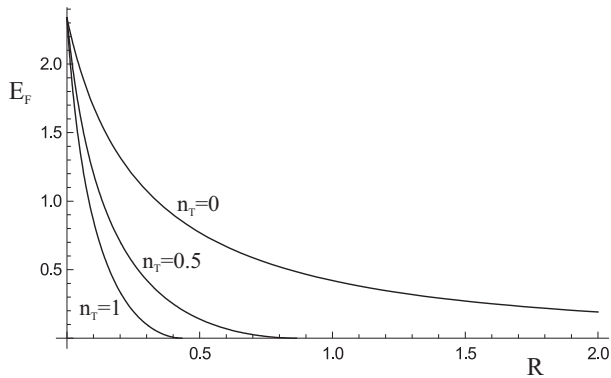}\\
  \caption{Entanglement of the asymptotic state as function of the ratio $R=\lambda/\kappa$ between the reservoir's coupling constants for some values of the number of thermal photons, $n_{T}$, with $r=1$ and arbitrary $\phi$. While the entanglement is always positive for $n_{T}=0$, if $n_{T}>0$ there is always some value of $R$ above which the entanglement is zero. 
 }\label{enform}
\end{figure}

Since the asymptotic states here obtained are also symmetric, we can
apply Eq.~\eqref{E_fRigol} to calculate their entanglement of
formation. In Fig.~\ref{enform}, we show their values as functions
of $R$ for some fixed values of $n_T$. For null temperature, when
there is no thermal photon, the entanglement is positive for any
rate. However, for each positive temperature, there is a maximal
rate above which the state is deep separable (as the coupling to the
natural reservoir grows, thermal effects are more sensible)
corresponding to the regime where the system exhibit SDE.

\subsection{Non-asymptotic dynamics}\label{Sub:Nonasymp}

We will discuss here in more detail the system intermediate
dynamics. This task is facilitated recognizing, by
Eqs.~\eqref{CMDyn}, that the dynamics will always describe a
straight line in the space of the parameters of the CM
($n_{1},n_{2}$, etc.)\footnote{Although the trajectories in the
space of density matrices itself will not be straight lines.}, for
every initial GS, and that the set of separable states is convex for
these parameters also. In other words, the trajectories of the CM
are given by straight line segments exponentially approaching the
asymptotic CM, $V_{\hat{\rho}_f}$. From now on we set $\phi=0$, for
simplicity in the analysis.

\begin{figure}[htp]
  \centering
  \includegraphics[width=8 cm]{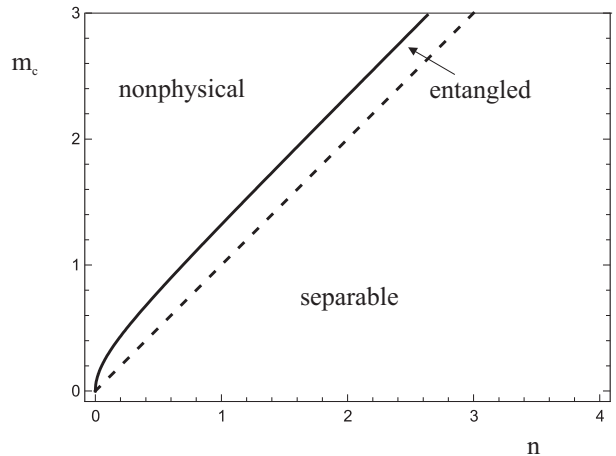}\\
  \caption{Representation of the CM's with $n_{1}=n_{2}=n$, $m_{c}\geq0$ and all other elements equal to zero. The region where these parameters represents physical CM's is divided by a straight line, above which all states are entangled, while below they are separable. Both regions are convex in the parameters.}\label{estab}
\end{figure}

\begin{figure}[htp]
  \centering
  \includegraphics[width=8 cm]{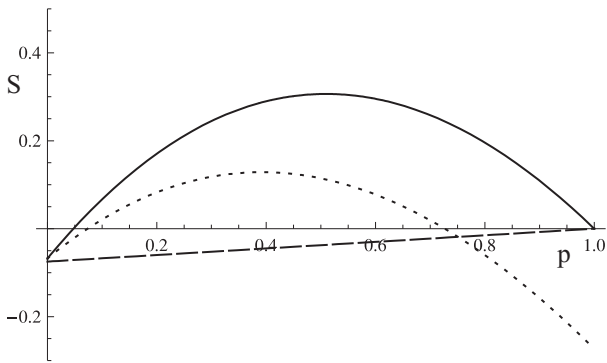}\\
  \caption{Evolution of the Simon  function $S$, parametrized by $p(t)$ for three distinct situations. The continuous line corresponds to an initially entangled state, which enters in the separable region at finite time and converges to an asymptotic state in the frontier. The dotted line is for the same initial state, but the reservoir is such that its corresponding asymptotic state is entangled, with the system losing all its entanglement to recover it later. Finally, the dashed line corresponds to a situation of asymptotic death of entanglement.}\label{evol}
\end{figure}

We begin with the situation where the reservoir parameters satisfy
Eq.~\eqref{eqcurva} so that the corresponding asymptotic state rests
in the border between the separable and entangled sets. If we
restrict our attention to initial states with CM, where
$n_{1}=n_{2}=n$, $m_{c}$ is real and all other elements are null,
Eqs.~\eqref{CMDyn} keep the dynamics inside this same subset. In
Fig.~\ref{estab}, we plot a diagram representing the values of
$n,m_{c}$ such that the corresponding CM is nonphysical, separable
or entangled.
The frontier between separable and entangled states is given by the
straight line $n=m_{c}$, so the entangled region is convex in these
parameters, and by our choice an asymptotic state is defined by the
dynamics somewhere in this line. A simple picture of the dynamics
can be given for these states, using the fact that the trajectory in
the $n \times m_{c}$ space will be a straight line, from the values
$(n_{0},m_{c,0})$ of the initial state to the asymptotic values
$(n_{f},m_{c,f})$, defined by reservoir parameters $r$ and $R$ [with
$n_{T}$ given by Eq.~\eqref{eqcurva}]. If these parameters are such
that $\de{n_{0},m_{c,0}}$ represents an entangled state, since
$\de{n_{f},m_{c,f}}$ belongs to the line $n=m_{c}$, the entire
trajectory will be on the entangled-state region, by convexity, \ie
entanglement will vanish asymptotically. To illustrate this, we plot
in Fig.~\ref{evol} (dashed line) the evolution of the $S$ function,
parametrized by $p(t)=1-\exp{[-2(\kappa+\lambda)t]}$, for an initial
state with $n_{0}=1$ and $m_{c,0}=1.0125$ and an asymptotic state
with $n_{f}=m_{c,f}=1$. Its value is initially negative, since the
initial state is entangled, and remains negative for all times.

\begin{figure}[htp]
  \centering
  \includegraphics[width=8 cm]{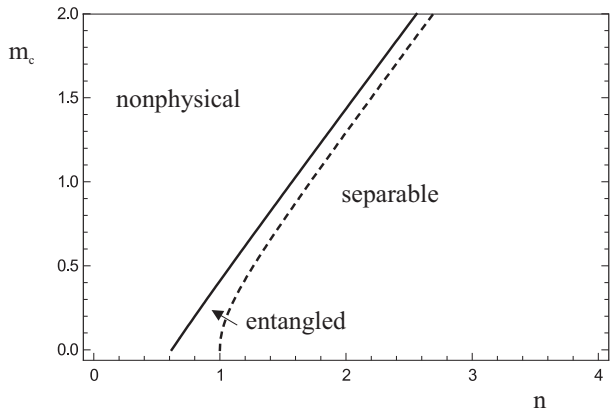}\\
  \caption{Representation of the CM's with $n_{1}=n_{2}=n$, $m_{1}=m_{2}=m\geq0$, $m_{c}=1$ and all other elements equal to zero. The region where these parameters represents physical CM's is divided by the dashed curve, above which all states are entangled, while below they are separable. Now the entangled region is not convex in the parameters.}\label{estac}
\end{figure}

For  examples of entanglement sudden death, consider the set of
states with $n_{1}=n_{2}=n$, $m_{c}=1$ and $m_{1}=m_{2}=m$ a real
number. In Fig.~\ref{estac}, we exhibit a diagram analogous to the
one in Fig.~\ref{estab}, and we see that the subset of entangled
states is not convex anymore. If the asymptotic state also has
$m_{c}=1$, the trajectory of the state CM can be described in this
diagram and will be again a straight line. Since the entangled
region is not convex, we may have initially entangled states that
will lose all entanglement at finite time, even if the asymptotic
state is in the frontier between the regions.
For instance, we can take parameters for the reservoir such that
$n_{f}=m_{c,f}=1$ and an initial CM that has elements $n_{0}=1.2$,
$m_{0}=0.5$, and $m_{c,0}=1$. In Fig.~\ref{evol} (continuous line),
we plot, as before, the function $S$ of the evolved state. We see
that it is initially negative, representing an entangled state, it
became positive at finite time, so the state has no entanglement,
and it remains strictly positive until it vanishes for $p=1$ (or
$t\rightarrow\infty$), because the state converges to a point in the
frontier.

If we perturb a little the reservoir parameters so that this
asymptotic state becomes entangled, and taking the same initial
state, the curve in the figure will be slightly distorted but now
will cross the $p$ axis twice, meaning that the state loses all
entanglement suddenly but recovers it a later time, remaining
entangled for the rest of the dynamics, a possibility mentioned
before (see the dotted line of Fig.~\ref{evol}, with the parameters
$n_{f}=0.95$ and $m_{c,f}=1$ for the asymptotic state). This is a
consequence of the fact that the set of entangled states is not
convex on these parameters. Cycles of birth and death are not
allowed, since a straight line can cross the convex set of separable
states only once.

\begin{figure}
 \includegraphics[width=8 cm]{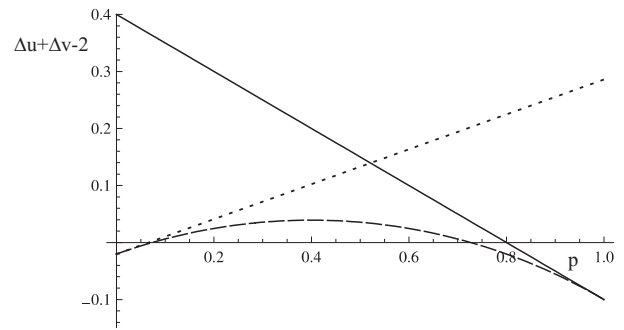}\\
  \caption{Evolution, parametrized by $p$, for the sum of variances of pairs of EPR quadratures, subtracted by $2$, so that a negative value indicates an entangled state. The continuous, dotted, and dashed curves are, respectively, for the optimum pairs of the final state, the initial state, and the state in each instant of time. We see that the common reservoir enlarges the optimum pair of the initial state to squeeze the pair of the asymptotic state (which is initially large). And there is a period of time where the quadratures with the least sum of variances are not squeezed enough to entangle the states.}\label{evolucaoquadraturas}
\end{figure}

This feature can be understood, from a more physical point of view,
considering the entanglement from the perspective of squeezed
EPR-like quadratures. Again, the EPR-like operators with least sum
of variances values (i.e., the optimum pair for this state) can be
found and read $\hat{X}_{1,r'}-\hat{X}_{2,r'}$ and
$\hat{P}_{1,r'}+\hat{P}_{2,'}$ with $\hat{X}_{i,r'}=r'\hat{x}_{i}$,
$\hat{P}_{i,r'}=\hat{p}_{i}/r'$ and $r'=\sqrt{(n-m-1/2)/(n+m-1/2)}$.
But these are not the quadratures ``chosen by'' the common reservoir
to squeeze (being those with $r'=1$, since we had set $\phi=0$). So,
on its way to squeezing its favorites, it enlarges the ones from the
initial state, such that there is a period of time when no pair of
EPR-like quadratures at all are squeezed enough to entangle the
modes (see Fig. \ref{evolucaoquadraturas}).

It remains then to explore the intermediate dynamics for the
situation where the asymptotic state is in the interior of the
separables, but this is also easily inferred from the straight line
trajectories exhibited by the system: every initial separable state
will remain separable for all times, and every initial entangled
state will lose its entanglement in finite time.


\subsection{Asymmetric engineered reservoir}

We have also considered the other extreme case where only one type
of atom enters in the cavity, say of type $1$, so that
$\kappa_{1}>0$ and $\kappa_{2}=0$ (but again
$\lambda_{1}=\lambda_{2}=\lambda$). Here, the second momenta
equations of motion read:
\begin{subequations}
\begin{eqnarray}
 \dot{n_{1}}&=& -2(A^2 \kappa+ \lambda) n_1  + (A B \kappa)m_c  + \nonumber \\
&& + (A B^* \kappa)m_c^*  + 2 \lambda n_T, \\
\dot{n_{2}}&=& -2(\lambda - |B|^2 \kappa) n_1 - m_c (A B
\kappa) - \nonumber \\
&& - (A B^* \kappa)m_c^*  + 2 \lambda n_T + 2 \kappa |B|^2, \\
\dot{m_1} &=& -2 (A^2\kappa + \lambda) m_1  -(2 A B^*
\kappa)m_s, \\
\dot{m_2} &=& -2 (\lambda - |B|^2\kappa)m_2+ (2AB^* \kappa)m_s^*, \\
\dot{m_c} &=& - (\kappa + 2 \lambda)m_c  - (A B^* \kappa)n_1  + \nonumber \\
 &&+  (A B^* \kappa)n_2 + A B^* \kappa, \\
\dot{m_s} &=& - (\kappa + 2\lambda)m_s + (A B \kappa)m_1  -   (A B^*
\kappa)m_2.
\end{eqnarray}
\end{subequations}

\begin{figure}[htp]
  \centering
  \includegraphics[width=8 cm]{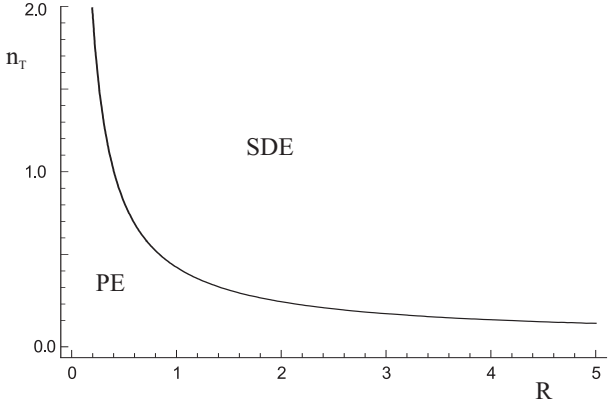}\\
  \caption{Phase diagram for the asymmetric reservoir, exhibiting the same qualitative behavior as in the symmetric case.}\label{assimetrico}
\end{figure}

Despite the asymmetry of the master equation here, the asymptotic
states have the same qualitative behavior as in the previous case:
for null temperature it will be entangled for any value of
$R=\kappa_{1}/\lambda$ but for finite temperature it is separable
for high enough $R$. In Fig.~\ref{assimetrico}, we plot the diagram
analogous to the one in Fig.~\ref{curva}, determining the dynamical
phases for the entanglement dynamics, also separated by a curve with
a similar shape.

\begin{figure}[htp]
  \centering
  \includegraphics[width=8 cm]{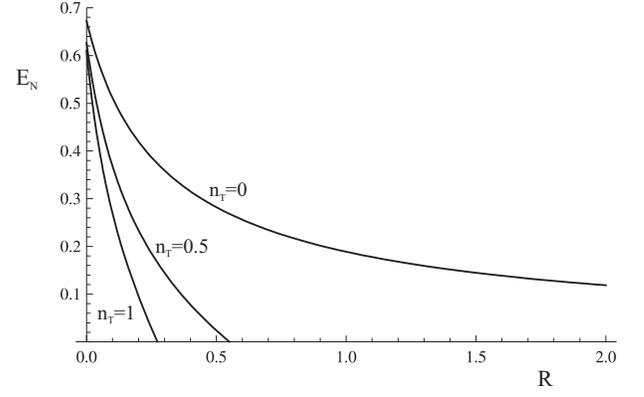}\\
  \caption{Log-negativity of the asymptotic state for the values $r=1$, and some values of $n_{T}$.}\label{assimetrico2}
\end{figure}

The asymptotic states entanglement, using now the log-negativity for
asymmetric GS~\cite{negcv}, are also shown (Fig.~\ref{assimetrico2})
for distinct values of $n_{T}$, as function of $R$, and the same
interpretation applies here.

Analogous results are valid also when $\kappa_{2}>0$ but
$\kappa_{1}\neq\kappa_{2}$. So, as far as asymptotic issues are
concerned, the system presents the same general behavior for both
symmetric and asymmetric reservoirs. However, the non-asymptotic
analysis made before for the symmetric reservoir becomes much more
involved for the asymmetric one, and will be subject of future work.

\section{Experimental Proposal}\label{sec:ExpProp}

The dynamics studied here is simpler than that in Ref.~\cite{PR}
(\eg being Markovian), yet it is much more controllable and with
clear experimental motivation. In this section, we discuss in more
detail how to test these predictions in the laboratory.

Considering the state evolution in
the laser reference frame, 
 Eqs.~\eqref{CMDyn} will be replaced by
\begin{subequations} \label{CMDynCG}
\begin{eqnarray}
 \dot{n_{j}}&=&-2(\kappa+\lambda)n_{j}+2\kappa|B|^{2}+2\lambda n_{T},\\
 \dot{m_{j}}&=&-2(\kappa+\lambda)m_{j} + (-1)^{j+1}2idm_j,\\
 \dot{m_c}&=&-2(\kappa+\lambda)m_c+2\kappa AB^{*}, \\
 \dot{m_s}&=&-2(\kappa+\lambda)m_s + 2idm_s,
\end{eqnarray}
\end{subequations}
where only the equations for $m_j$ and $m_s$ have changed.
If we consider the initial GS where $m_{j}=m_{s}=0$, as well as null
amplitudes, the evolution will be the same as that for the
interaction picture, i.e., given by Eq.\eqref{master}, and the
coarse graining in time will not affect these states throughout the
evolution. For other initial states, the coarse graining will be a
source of decoherence, and the state may even lose the Gaussian
character along the evolution (like a one-mode coherent state, with
non-vanishing amplitude, averaged over its phase).

Our proposal starts then from the preparation of a highly entangled
state using the common symmetric reservoir~\cite{davidov} for a
total time much shorter than $\lambda^{-1}$, and setting $R\ll 1$ in
order to make thermal effects negligible. After the preparation
time, the parameter $R$ is set to values of the order
$R\gtrapprox1$, by changing the flux of atoms through the cavity
\footnote{Noting that changing the driving field amplitude alters
also the squeezing parameter, and that interrupting the flux
corresponds to $R\rightarrow\infty$.}. The system is allowed to
evolve until some time $t$ (of the order $\lambda^{-1}$) and the
entanglement evolution can be studied, for example, by using the
method proposed in Ref.~\cite{marcelin}, to reconstruct the two-mode
state. Naturally, one could also try to measure some entanglement
witness to simplify the experimental procedure, by using the
previous knowledge of the field to reduce the amount of experimental
data, but still characterize its entanglement.

\begin{figure}[htp]
  \centering
  \includegraphics[width=8 cm]{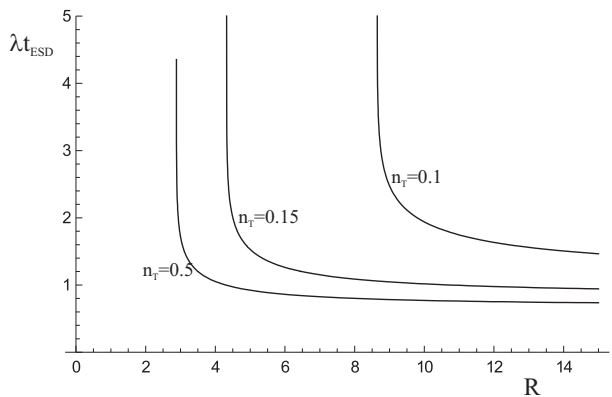}\\
  \caption{Sudden death time as function of $R$, for some values of $n_{T}$. It is infinite for small $R$, corresponding to the persistent entanglement region, but after the threshold, it rapidly falls to a value near to unity.}\label{tempoms}
\end{figure}

For this setting, \emph{i.e.}, assuming an initial two-mode squeezed
state, the times where entanglement sudden death (ESD) takes place
can be obtained explicitly and reads:
\begin{equation}
\lambda t_{ESD}=\frac{R}{2(1+R)}\ln{(1-p_{ESD})^{-1}},
\end{equation}
where
\begin{equation}
p_{ESD}=\frac{(1+R)(B^{2}-AB)}{(1+R)(B^{2}-AB)-B^{2}-n_{T}R+AB},
\end{equation}
and the parameter $R$ refers to the one in the second stage of the
procedure, after the initial state preparation. These times, in
units of $\lambda^{-1}$, are exhibited as function of $R$ in
Fig.~\ref{tempoms}, for some values of temperature. Naturally, they
are infinite for small $R$, in the region where persistent
entanglement takes place. After the threshold, it falls abruptly and
stabilizes in values close to unity, so the entanglement will
typically survive long enough for its (not so sudden) death to be
observed. It may also be interesting to study thermal effects like
the behavior of $\frac{d E_{F}}{dt}$ before the sudden death, which
is an indicator of the incidence of the dynamical trajectory of
$\hat{\rho}$ at the set of separable states, by controlling the
temperature.

\section{Robustness of the Scheme for Generating Entangled GS in a Cavity}

Since entanglement in GS is related to squeezing in the proper EPR
quadratures, while the natural thermal effect is to spread Wigner
functions, it is in order to ask about the sensibility of the
preparation scheme with natural dissipation, considering the regime
$R\ll1$ and $n_{T}\ll1$. Actually we have analyzed a slightly
different procedure than the original one by Pielawa \emph{et al.}
\cite{davidov}. There they propose to empty the modes $\hat{b}_{j}$
at turns, passing a stream of atoms of type $1$, say, until mode
$\hat{b}_{1}$ is sufficiently ``washed'', then repeating the
procedure with type $2$ atoms. Here we consider that both types of
atoms can pass through the cavity, not at the same time, but with
equal probability, so that we can apply the equations for the
symmetric reservoir to compute the system evolution.

\begin{figure}[htp]
  \centering
  \includegraphics[width=8 cm]{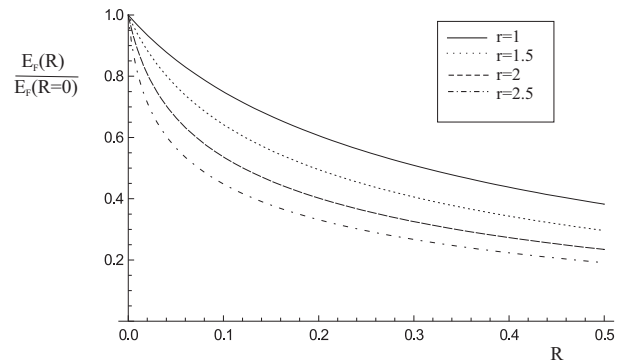}\\
  \caption{Entanglement of the asymptotic states as function of the ratio $R=\lambda/\kappa$ between the reservoir decay rates, for some values of squeezing parameter $r$, with $n_{T}=0.05$, normalized by its value at $R=0$. The continuous line is for $r=1$, dotted for $r=1.5$, dashed for $r=2$ and dotted-dashed for $r=2.5$. The entanglement is somewhat sensitive to dissipation, and is more sensitive the greater the squeezing parameter.}\label{analiseexp}
\end{figure}

Supposing that the two original modes, $\hat{a}_i$, are initially
empty (and remembering that the number of photons can be lower than
the number of thermal photons if one passes a suitable stream of
atoms to ``steal'' photons from the mode), the evolution of the
covariance matrix of the system will be then given with the
following non-zero entries:
$n_{1}=n_{2}=\frac{|B|^{2}+n_{T}R}{1+R}p(t)$ and
$m_{c}=\frac{AB}{1+R}p(t)$, where again
$p(t)=(1-\exp{[-2(\kappa+\lambda)t]})$. Assuming also that the
duration of experiment $t$ is large enough so that $p\approx 1$
(which can be achieved with $t\gg \kappa^{-1}$ but still
$t\ll\lambda^{-1}$) or, in other words, that the system is
essentially in the asymptotic state of the procedure, we plot in
Fig.~\ref{analiseexp} the entanglement of formation as a function of
$R$, for $n_{T}=0.05$ (value attained in the recent experiment
reported in Ref.~\cite{expharoche}) for distinct values of $r$.
The graphs are normalized by the value of the entanglement of
formation with $R=0$, which would be obtained if there were no
dissipation. We see that the entanglement is somewhat sensitive to
the dissipation, even at such a low temperature, and can be lowered
by half, for $R=0.1$, \ie, when the engineered reservoir rate is ten
times greater then the dissipation rate. Also, the greater the
squeezing parameter, the sensitive entanglement becomes to
dissipation, as expected. Though, if $\kappa$ is above two orders of
magnitude greater than $\lambda$, the entanglement does not appear
to be significantly changed. For a squeezing parameter of $r=1$, a
value of $R\approx 10^{-2}$ and a waiting time of about $3
\kappa^{-1}$ would suffice to obtain an amount of entanglement
greater than $90\%$ of the pure ideal state (with $R=0$ and infinite
waiting time).

\section{Conclusion}

An engineered reservoir can be used to create entangled states of
two field modes in a cavity~\cite{davidov}. We here show that it can
also be used, together with the natural thermal reservoir, to depict
asymptotic entanglement phase diagrams~\cite{PR}. Taking advantage
of the fact that the specific reservoirs used preserve Gaussianity,
we could make all the discussion of entanglement in terms of
covariance matrices. We showed that only two behaviors are allowed:
asymptotic entanglement or entanglement ``sudden death''. The line
between both phases allows the coexistence of both behaviors.

Besides asymptotic studies, the symmetric engineered reservoir was
also studied in detail, with examples of all possible entanglement
fates exhibited.

This study can be viewed as an experimental proposal for drawing
entanglement phase diagrams and for monitoring the ``transition''.
Another byproduct is the study of the robustness of the proposal for
entanglement generation with respect to the natural thermal
environment. An interesting question raised is what would be the
best setup for such experimental drawing of asymptotic entanglement
phase diagrams.

The authors thank S. Pielawa, M. F. Santos and J. P. Paz for helpful
comments and suggestions. We are grateful for the support of
Brazilian agencies CNPq and Fapemig. This work is part of Brazilian
National Institute for Science and Technology on Quantum
Information. R.C.D. was supported also by CAPES, Proc.: BEX
0124/10-9.

\end{document}